# ON THE RADIATION OF AN ARBITRARILY MOVING CONSTANT MAGNETIC MOMENT

B. A. Zon[1], V. E. Chernov[1], and M. Ya. Amusia[2,3]

Formulas determining the electromagnetic field emitted by a uniformly accelerated moving free charge, including the Larmor formula for the intensity of this radiation, are well known [1]. It would be quite natural to expect such radiation also to exist for the magnetic moment. Such a possibility would be of compelling interest for astrophysics since a constant magnetic moment associated with the spin of the electron exists in the hydrogen atom – the most widely distributed element in the Universe. Weak electromagnetic fields emitted by the hydrogen atom can be used in the diagnostics of gravitational fields, including those associated with fluctuations of the dark matter density [2–4]. This problem acquires additional urgency in light of the substantial progress that has been achieved in the development of superconducting-nanowire-based single-photon detectors [5–7].

In this note, we show, however, that such radiation is absent, not only for uniformly accelerated motion of a magnetic moment, which is characteristic of the motion of a hydrogen atom in a uniform gravitational field, but also of its motion according to an arbitrary law. Let us turn now to the calculations. We shall describe the radiation of electromagnetic waves with the help of the Hertz vector $\mathbf{Z}$. For a variable magnetic moment $\mathbf{M}$ the Hertz vector far from the region where the magnetic moment is located (in the wave zone), is given by the expression (see [8], p. 545, for example)

$$\mathbf{Z}(\mathbf{r}, t) = [\mathbf{M}(t - \tau) \times \mathbf{n}] / r. \tag{1}$$

Here $\mathbf{n} = \mathbf{r}/r$ is the unit vector directed from the emitting region, which we will assume to be located near the coordinate origin, to the observation point $\mathbf{r}$, $\tau = r/c$ is the delay time, and $c$ is the speed of light. Now let the constant magnetic moment $\mathbf{M}$ move along some trajectory $\mathbf{r}_0(t)$. We thus obtain for the Hertz vector, in analogy with Eq. (1), the following formulas:

$$\mathbf{Z}(\mathbf{r}, t) = [\mathbf{M} \times \mathbf{N}] / |\mathbf{r} - \mathbf{r}_0(t - T)|, \tag{2}$$

$$\mathbf{N} = (\mathbf{r} - \mathbf{r}_0(t - T)) / |\mathbf{r} - \mathbf{r}_0(t - T)|,$$

$$T = |\mathbf{r} - \mathbf{r}_0(t - T)| / c.$$

As can be seen, the delay time $T$ is now determined implicitly as the solution of some equation. For large distances $r$ from the magnetic moment to the observation point, significantly exceeding the finite values $\mathbf{r}_0(t - T)$, it follows from formula (2) that

$$\mathbf{Z}(\mathbf{r}, t) \sim 1 / r, \tag{3}$$

as it should be for radiation of electromagnetic waves. However, for the vector potential given in terms of the Hertz vector by the expression

$$A = (1/c)\, \partial \mathbf{Z}/\partial t, \tag{4}$$

a relation analogous to formula (3) no longer applies. Indeed, it is not difficult to see that differentiating Eqs. (2) with respect to time yields the dependence

$$A \sim 1/r^2, \tag{5}$$

but such a rapid falloff of the field at large distances from the source no longer corresponds to radiation of electromagnetic waves. Thus, a constant magnetic moment moving along an arbitrary trajectory does not radiate electromagnetic waves.

The obtained result admits a quite clear physical interpretation. Radiated energy can arise, in principle, only from the kinetic energy of a particle possessing a magnetic moment, in our case – from the kinetic energy of a moving hydrogen atom. However, the kinetic energy of the hydrogen atom in no way enters into the Maxwell equations describing electromagnetic radiation. Therefore, the transformation of kinetic energy into electromagnetic energy does not take place.

It is also important to note that the above consideration assumes conservation during the motion of the hydrogen atom of the constant direction of its spin magnetic moment. Such an assumption is completely natural for quantum particles since the direction of the electron spin is an integral of the motion (neglecting the hyperfine interaction). The solution of the problem of radiation of a classical particle possessing a magnetic moment and rotating about an axis that does not coincide with the direction of the magnetic moment is given in [8].

This work was performed with partial financial support of the Russian Foundation for Basic Research and the Czech Scientific Foundation within the scope of Scientific Project No. 19-52-26006.

**REFERENCES**

1. L. D. Landau and E. M. Lifshitz, The Classical Theory of Fields, Butterworth-Heinemann, London (1975).
2. A. N. Taylor and M. Rowan-Robinson, Nature, **359**, 396 (1992). DOI: 10.1038/359396a0.
3. S. Dodelson, Modern Cosmology, Academic Press, Cambridge, MA (2003).
4. A. Maeder and V. G. Gueorguiev, Phys. Dark Universe, **25**, 100315 (2019). DOI: 10.1016/j.dark.2019.100315.
5. N. Calandri, Q-Y. Zhao, D. Zhu, *et al.*, Appl. Phys. Lett., **109**, 152601 (2016). DOI: 10.1063/1.4963158.
6. Q-Y. Zhao, D. Zhu, N. Calandri, *et al.*, Nature Photonics, **11**, 247 (2017). DOI: 10.1038/nphoton. 2017.35.
7. B. Korzh, Q-Y. Zhao, J. Allmaras, *et al.*, Nature Photonics, **14**, 250 (2020). DOI: 10.1038/s41566-020-0589-x.
8. V. V. Batygin and I. N. Toptygin, Collection of Problems on Electrodynamics [in Russian], Scientific Publishing Center: Regular and Chaotic Dynamics Publishing House, Moscow (2002).

[1]Voronezh State University, Voronezh, Russia, e-mail: zon@niif.vsu.ru; chernov@niif.vsu.ru; [2]A. F. Ioffe Physical-Technical Institute of the Russian Academy of

Sciences, Saint Petersburg, Russia, e-mail: Miron.Amusia@mail.huji.ac.il; [3] G. Racah Institute of Physics of the Hebrew University, Jerusalem, Israel.